# Characterization of 250 MeV protons from Varian ProBeam pencil beam scanning system for FLASH radiation therapy

Serdar Charyyev[1,*], Chih-Wei Chang[2], Mingyao Zhu[2], Liyong Lin[2], Katja Langen[2], Anees Dhabaan[2]

[1] Department of Radiation Oncology, Stanford University, Palo Alto, CA 94304, USA

[2] Department of Radiation Oncology and Winship Cancer Institute, Emory University, Atlanta, GA 30322, USA

Running title: **Characterization of 250 MeV proton PBS for FLASH**

The authors have no conflicts to disclose.

**Abstract**

**Purpose:** Recently, 'shoot-through' proton FLASH radiation therapy (RT) has been proposed where the highest energy is extracted from the cyclotron to maximize the dose rate (DR). Even though our proton pencil beam scanning (PBS) system, Varian ProBeam, is capable of delivering 250 MeV (the highest energy), it is not typical to use 250 MeV protons for routine clinical treatments and as such 250 MeV may not have been characterized in the commissioning. In this study, we aim to characterize 250 MeV protons from Varian ProBeam system for FLASH RT as well as assess the ability of clinical monitoring ionization chamber (MIC) for FLASH-readiness.

**Materials and Methods:** We measured data needed for beam commissioning: integral depth dose (IDD) curve, spot sigma, and absolute dose calibration. To evaluate MIC, we measured output as a function of beam current. To characterize a 250 MeV FLASH beam, we measured: (1) central axis DR as a function of current and spot spacing and arrangement, (2) for a fixed spot spacing, the maximum field size that still achieves FLASH DR (i.e., > 40 Gy/s), (3) DR reproducibility. All FLASH DR measurements were performed using ion chamber for the absolute dose and irradiation times were obtained from log files of ProBeam. We independently verified dose measurements using EBT-XD films and irradiation times using a fast, pixelated spectral detector.

**Results:** R90 and R80 from IDD were 37.58 and 37.69 cm, and spot sigma at isocenter were $\sigma_x$=3.336 and $\sigma_y$=3.332 mm, respectively. The absolute dose output was measured as 0.377 GyE*mm$^2$/MU for the commissioning conditions. Output was stable for beam currents up to 15 nA, and it gradually increased to 12-fold for 115 nA. DR depended on beam current, spot spacing and arrangement and could be reproduced within 4.2% variations.

**Conclusion:** Even though FLASH was achieved and the largest field size that delivers FLASH DR was determined as 35x35 mm$^2$, current MIC has DR dependence and users should measure DR each time for their FLASH applications.

## 1. Introduction

A sequence of research (1-4) has shown that ultra-high DR (>40 Gy/s, FLASH) has a sparing effect on normal tissue due to 'FLASH effect'. The seminal work by Favaudon et al. (1) demonstrated normal tissue sparing in nude mice in which they used low-energy electrons to irradiate their target. The same FLASH irradiator was used to treat the first patient with FLASH radiation therapy (RT), a cutaneous lymphoma patient (5). Even though they are suitable for skin cancer, electrons are not penetrative enough to produce useful beam for deep-seated tumors.

There are investigations that have shown FLASH effect with x-rays in kV range (6-8) and efforts to achieve FLASH effect with MV x-rays (9). Regardless, an attractive and readily available option to deliver a FLASH beam is with cyclotron-based protons, specifically pencil beam scanning (PBS), since PBS can be better controlled in terms of position and intensity (10). For these systems, proton therapy is delivered at a clinical dose rate (DR) of a few Gy/minute (~1-2 Gy/min). Switching to 'FLASH mode' necessitates increasing the DR by orders of magnitude. This is a challenge for treatment delivery systems, especially at low energies, because energy switching process causes beam losses that limits beam currents that reach the treatment rooms.

Recently, techniques where proton beams shoot through the patient from different angles and irradiate the tumor with the plateau region, namely 'shoot-through' proton FLASH RT, have been proposed (11, 12). Even though this means that the tissue at the distal end of the tumor will receive a higher dose relative to conventional proton RT, the proposal of shoot-through FLASH emerged out of an interest in maximizing the DR. To deliver the proton beam at FLASH DR on a cyclotron-based proton therapy system, the inefficient beam degrader and energy selection system need to be removed, such that the beam with the highest energy can be extracted from the cyclotron (13).

Even though our PBS system, Varian ProBeam, is capable of delivering 250 MeV (the highest energy) proton beam, it is not typical to use 250 MeV protons for routine clinical treatments and as such 250 MeV may not have been characterized in the commissioning (14).

Ionization chambers are routinely used in RT to monitor absolute dose and are integrated in the delivery and monitoring system of our proton therapy system, and we rely on them to switch off the beam in timely manner. Therefore, their ability to work at a higher DR needs to be assessed before such systems can be released as 'FLASH-ready' for treatments.

In this study, we aim to characterize 250 MeV protons from a Varian ProBeam PBS system (Varian Medical Systems, Palo Alto) for FLASH RT as well as assess the ability of currently installed monitoring ionization chamber (MIC) for FLASH-readiness. We believe this is an important prerequisite for the safe and reliable delivery of FLASH with PBS. By doing this characterization, we can set up a beam model in our treatment planning system (TPS) and also have a set of reference values for subsequent quality assurance (QA) for that particular energy. This information is also essential for Monte Carlo (MC) applications, where beam characteristics are needed to set up the phase space parameters and run the simulations in order to avoid modeling the whole beamline (15, 29).

To date, two published studies have achieved FLASH DR with 250 MeV protons from Varian's ProBeam system (16, 17), but none reported the characteristics of the beam. This is necessary to provide guidance for prospective animal studies or clinical trials that intend to use the ProBeam system.

## 2. Materials and Methods

### 2.1. Measurement Data

In the context of this study, we measured data needed for beam commissioning as recommended per our TPS, RayStation: integrated depth dose (IDD) curve, spot sigma and virtual source position, and absolute dose calibration. To test the capabilities of MIC, we measured output as a function of beam current. To characterize 250 MeV for FLASH, we measured: (1) central axis (CAX) DR as a function of current and spot spacing and arrangement, (2) for a fixed spot spacing, the largest field size that still achieves FLASH DR (i.e., > 40 Gy/s), (3) DR reproducibility. All measurements with 250 MeV proton beam were performed in Varian's RaceHorse mode.

### 2.2. Spot sigma measurements

The spot profiles at six locations in air along the z-axis (400, 300, 200, 100, 0 and −100 mm; first four locations are upstream from the isocenter and the last location is downstream from the isocenter) were acquired for 250 MeV protons using a 2-dimensional scintillation detector Lynx (IBA Dosimetry, Schwarzenbruck, Germany) with 0.5-mm resolution. Spot sigmas in x-direction ($\sigma_x$) and y-direction ($\sigma_y$) were determined for each of the six locations by fitting the spot profile to a Gaussian distribution. The change in spot profile at different locations from the isocenter along the z-axis is used to determine phase space parameters of divergence and correlation coefficients (14) using RayStation (18).

### 2.3. Integrated depth dose measurements

The IDD curve was measured for monoenergetic 250 MeV proton beam incident on the CAX of a large-area detector, StingRay (IBA Dosimetry, Schwarzenbruck, Germany), with active scanning diameter of 12 cm. Integrated charge measurements for the entire Bragg peak (BP) were performed with DOSE-1 (IBA Dosimetry, Schwarzenbruck, Germany), a single channel, high precision, reference class electrometer by applying a bias of 500 V. The StingRay was scanned in a water tank with micro-levelling capabilities, Blue Phantom$^2$ (IBA Dosimetry, Schwarzenbruck, Germany). Because RaceHorse was commissioned at gantry angle of 0 and 90 degrees, IDD curve was measured at 90 degree gantry angle. IDD curve was shifted to account for the water-equivalent thickness (WET) of the pin distance, chamber buildup, tank wall thickness and an additional offset to avoid chamber damage, a total of 13.77 mm. Range from IDD was compared to continuous-slowing-down approximation (CSDA) range, as obtained from National Institute of Standards and Technology (NIST) data for protons (19).

### 2.4. Absolute dose calibration measurements

Absolute field output was measured in solid water, Gammex (Sun Nuclear Corp, Melbourne, Florida), with a 20-mm diameter parallel plate ionization chamber, PPC40 (IBA Dosimetry, Schwarzenbruck, Germany), and DOSE-1 electrometer at 2 cm depth using monoenergetic 10x10 cm$^2$ square field with 4 mm spot spacing (i.e., 25 spots in both x- and y-direction, with a total of 625 spots). Weight of each spot was 100 MU/spot to create a uniform lateral profile. The measured physical dose was scaled up by a factor of 1.1 to account for relative biological effectiveness (RBE) of protons. For commissioning purpose, we measured output with a beam current of 1 nA in source-to-axis distance (SAD) setup with gantry angle of 0 degrees. We then varied the beam current up to 115 nA to see variations of output with beam current. We also did measurements of output at different gantry angle, 90 degrees, and with source-to-surface distance (SSD) setup to see the dependence of the output to those variables.

### 2.5. Details of FLASH measurements

DR in this study is reported in terms of average DR, i.e., $\frac{delivered\ dose}{time\ of\ irradiation}$. Delivered dose is measured using PPC40 and DOSE-1 and times of irradiation were obtained from Varian log files. We independently verified dose measurements using EBT-XD radiochromic films. Verification measurements of the irradiation times were also performed using a fast, pixelated spectral detector AdvaPIX-TPX3 (TPX3) (20). DR dependence on beam current measurements were performed using a 35x35 mm$^2$ field size, with square spot arrangement and a 6 mm spot spacing to deliver 17.25 Gy at 10 cm depth. At 115 nA beam current, we measured DR dependence on spot spacing using 4, 6, 7, and 8 mm spot spacings and spot arrangement using a hexagonal spot arrangement. We modified field size on the fly in RaceHorse mode to determine the largest field size that still achieves FLASH DR at the beam current of 115 nA. For the DR reproducibility, we decreased the field size to 30x30 mm² and measured DR repeatedly ten times at the beam current of 150 nA.

## 3. Results

### 3.1. Spot sigma

In Figure 1 we present: (a) x- and y- spot profiles at isocenter (i.e., z=0) and corresponding fit to the Gaussian distribution, (b) same information as in (a) but y-axis is in log scale, (c) fitted Gaussian curves only, with $\sigma_x$=3.336 mm and $\sigma_y$=3.332 mm.

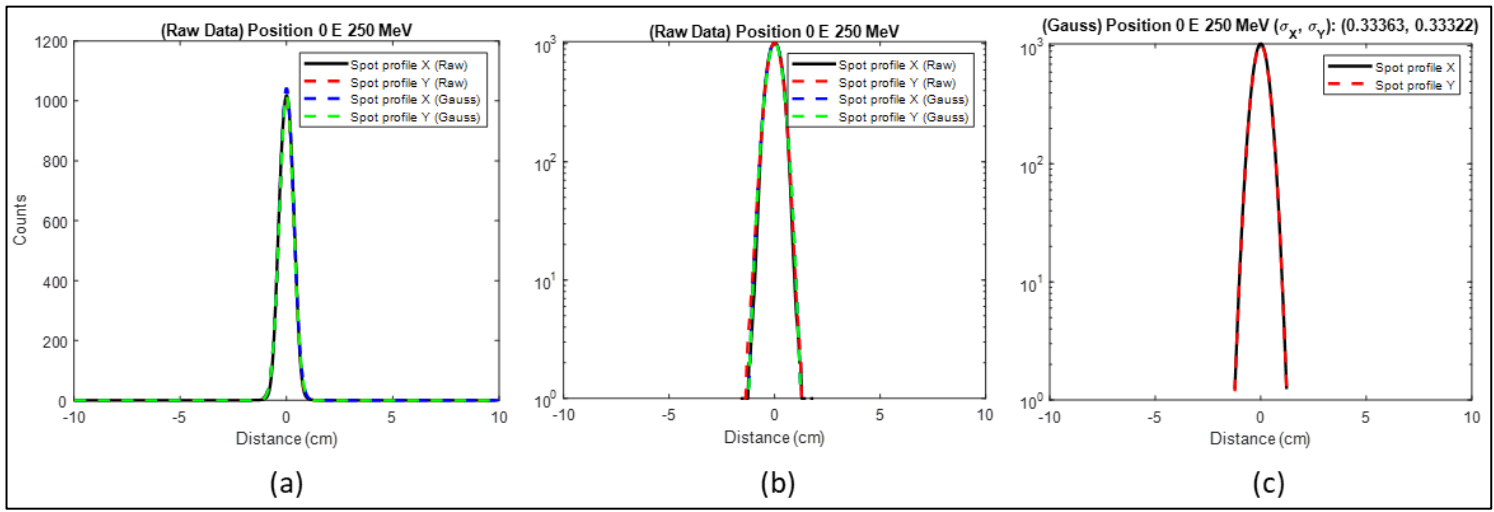

**Figure 1.** (a) x- and y- spot profiles at isocenter and corresponding fit to the Gaussian distribution, (b) same information as in (a) but y-axis is in log scale, (c) fitted Gaussian curves only (y-axis is in log scale).

Spot sigmas, $\sigma_x$ and $\sigma_y$, for all the measured six locations are tabulated in Table 1. From these measurements, we can determine the phase space parameters of divergence and correlation coefficient as 0.00154 rad and 0.6215.

**Table 1.** Spot sigmas, $\sigma_x$ and $\sigma_y$, for all of the measured six locations.

| Location (mm) | $\sigma_x$ (mm) | $\sigma_y$ (mm) |
|---|---|---|
| 400 | 3.084 | 2.8424 |
| 300 | 3.1367 | 3.0099 |
| 200 | 3.1754 | 3.1005 |
| 100 | 3.229 | 3.1969 |
| 0 | 3.3363 | 3.3322 |
| -100 | 3.3876 | 3.4459 |

### 3.2. Integrated depth dose

IDD curve measured with Stingray is shown in Figure 2. R90 and R80 were determined as 37.58 and 37.69 cm, respectively and CSDA range from NIST data is 37.94 cm.

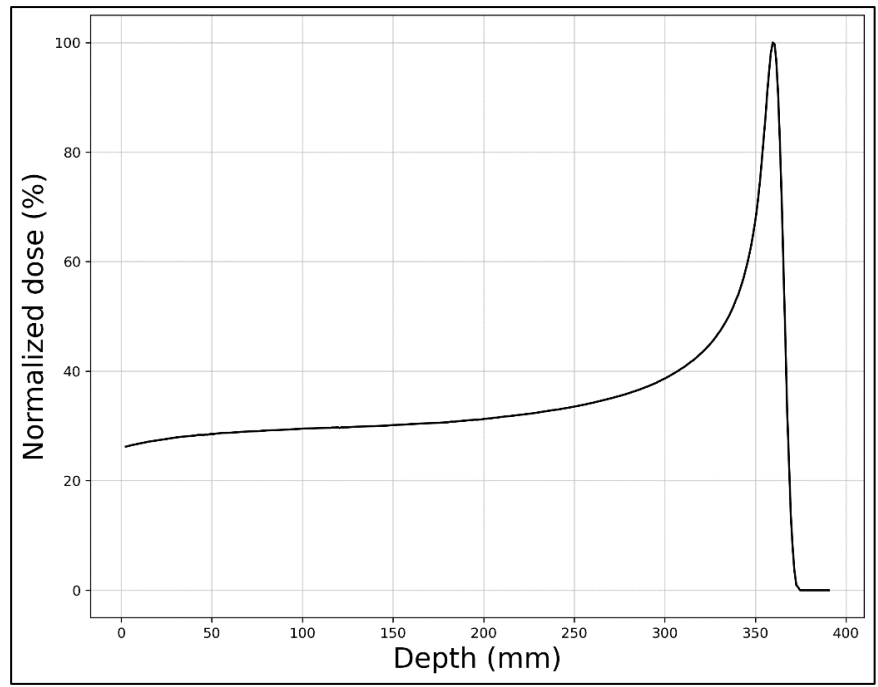

**Figure 2.** IDD curve for 250 MeV proton beam measured with Stingray.

### 3.3. Absolute dose calibration

The absolute dose output was measured as 0.377 GyE*mm$^2$/MU for the conditions specified in Section 2.4. Output is stable for beam currents up to 15 nA, after which it increases up to 4.794 GyE*mm$^2$/MU for beam current of 115 nA as can be seen from Figure 3. Output was measured to be independent of gantry angle (shown as red diamond symbol in Figure 3) and SSD/SAD setup (shown as black plus symbol in Figure 3) variations.

### 3.4. FLASH Measurements

As can be seen from Figure 4, DR increases with beam current (blue circles in Figure 4) for a 35x35 mm$^2$ field at a depth of 10 cm with ~0.3 Gy/s at 1 nA and ~40 Gy/s at 115 nA. Moreover, smaller spot spacings were delivered with a higher DR (black plus symbols in Figure 4) and hexagonal spot arrangement was delivered with slightly less DR (red diamond symbol in Figure 4) than a square arrangement with the same spot spacing. For 115 nA beam current, the largest field size that delivers FLASH DR was determined as 35x35 mm$^2$. A DR of 50 Gy/s and 56.6 Gy/s can be achieved for 30x30 mm$^2$ and 25x30 mm$^2$ field sizes, respectively, Figure 5a.

DR reproducibility for a 30x30 mm$^2$ field is illustrated in Figure 5b, as delivered with 150 nA beam current. The minimum, maximum and average DRs were of 62.4.8, 66.8 and 64.1 Gy/s, respectively and variability of up to 4.2% was observed. For these fields, a dose of 17.5 Gy was delivered at 10 cm depth as measured with PPC40. This was confirmed with EBT-XD measurement, as can be seen from Figure 6. Time of irradiation for one of the fields was 269 ms from Varian logs and confirmed as 268.9 ms with TPX3 detector.

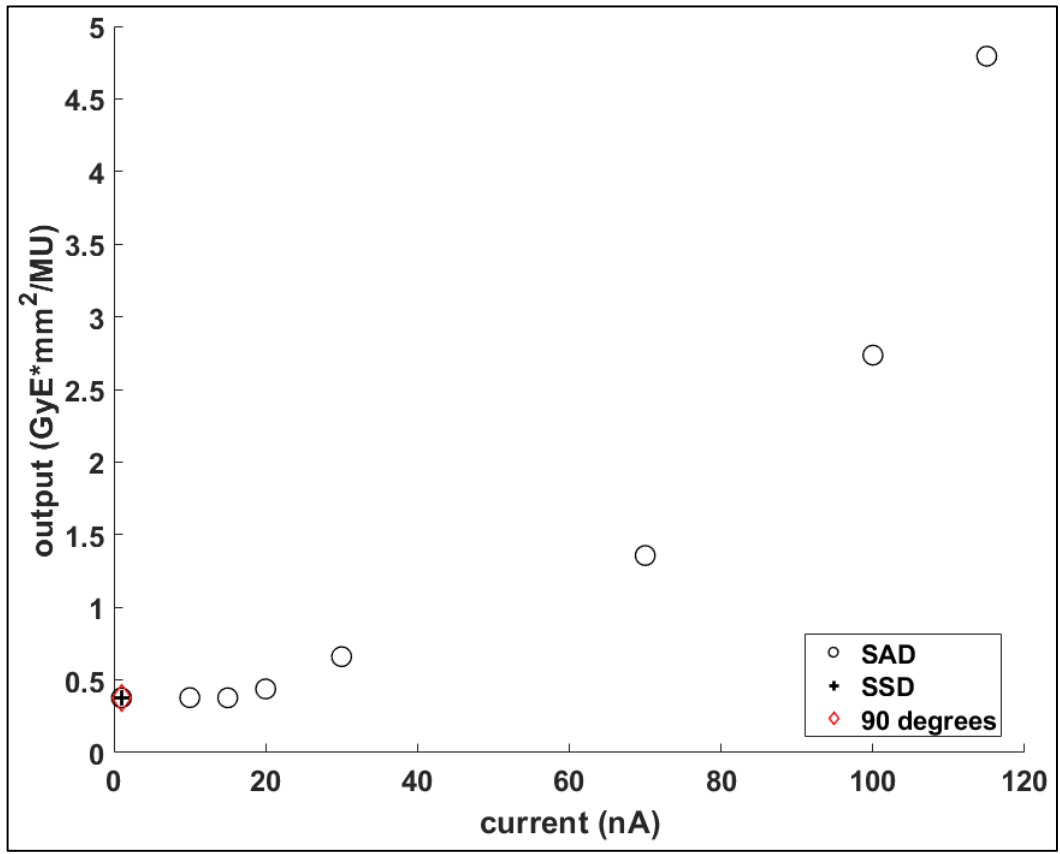

**Figure 3.** Output as a function of current in SAD setup (shown as black circles), SSD setup (shown as black plus symbol) and gantry angle (shown as red diamond symbol).

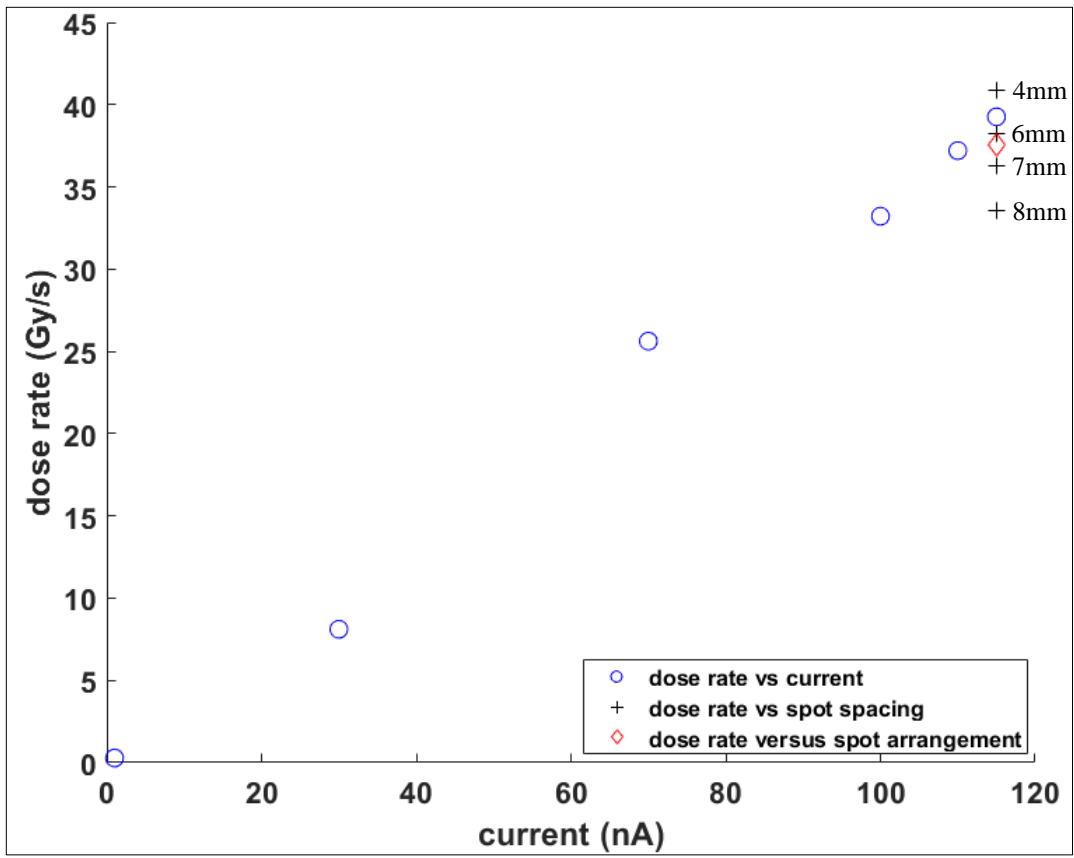

**Figure 4.** Dose rate as a function of beam current (blue circles), spot spacing (black plus symbol with corresponding spot spacing indicated on the figure) and spot arrangement (red diamond symbol for a hexagonal arrangement with 6 mm spot spacing).

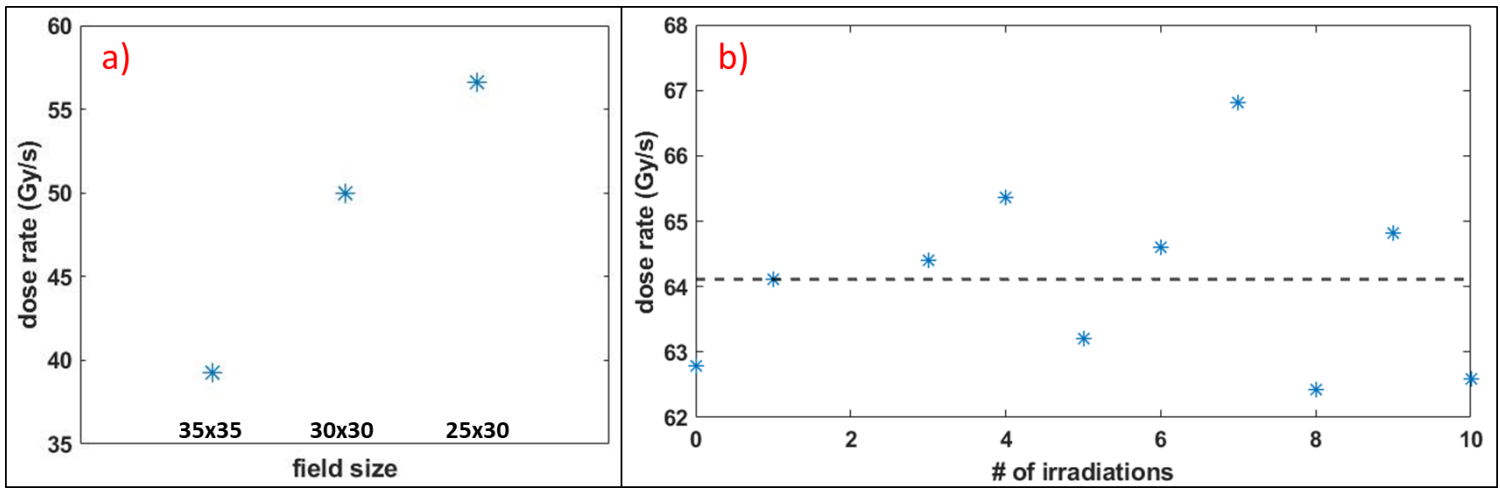

**Figure 5.** a) Dose rate as a function of field size at beam current 115nA. b) Dose rate reproducibility for a 30x30 mm$^2$ field delivered at 150 nA beam current for ten repeated irradiations with an average dose rate of 64.1 Gy/s, shown with a black dashed line.

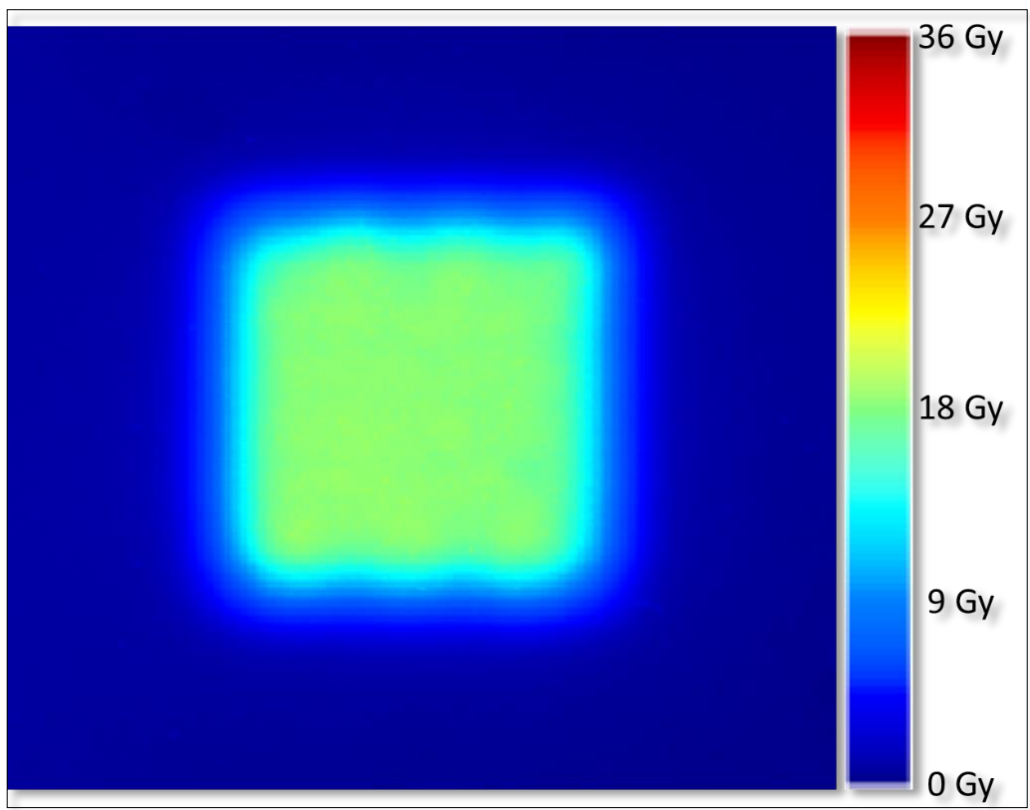

**Figure 6.** 2D dose distribution at 10 cm depth in solid water phantom for a 30x30 mm$^2$ field delivered at 150 nA beam current as measured with EBT-XD.

## 4. Discussion

Among the results presented in Section 3, several observations are noted and discussed in this section. First of all, output was found to be stable up to beam current of 15 nA, after which we have shown DR dependence of the MIC. Thus, the clinical MIC currently installed in ProBeam systems will not shut off the beam at the requested monitor units. This is likely due to higher ion recombination effects (21), which was demonstrated by Liszka et al. (22) and Yin et al. (23) to induce significant ion collection issues even at 13 Gy/s DR level. Until the design of the MIC is modified to account for this and subsequent replacement of MIC by the vendor, users should independently measure DR each time for their FLASH applications.

We have observed through reproducibility measurements that the DR can vary by about 3% for the repeated irradiations. This warrants caution. Because of the instabilities of the MIC to control the beam, secondary, possibly tertiary, confirmation measurements of the DR have to be performed. This uncertainty also needs to be taken into account when interpreting outcomes of studies with current clinical system.

DR for most FLASH studies is reported in terms of average DR, i.e., $\frac{delivered\ dose}{time\ of\ irradiation}$, which was the definition we adopted in this study. Even though there is a consensus that average DR is most relevant for FLASH effect (24-26), for proton PBS, DR can be defined, at each point in the field, as the sum of contributions from multiple spots (27, 28). In that sense, considering the subset of voxels in a field, the DR can be higher that the reported DR in this work. It is relevant to study DR in the context of sub-voxels, because the processes that are believed to be responsible for the FLASH effect happen at the cellular level.

## 5. Conclusion

In conclusion, we characterized 250 MeV protons and evaluated the reliability of MIC from a Varian ProBeam PBS system for FLASH RT. The values we provided will serve as reference values for quality checks. Moreover, the essential beam phase space parameters are provided to explore FLASH effects using proton MC. We investigated the clinical MIC for FLASH-readiness and have concluded that users should make independent and repeated measurements of DR until MIC design is changed to accurately monitor the FLASH beam.